\documentclass[twocolumn,prb,superscriptaddress,preprintnumbers ,showpacs,amsmath,amssymb]{revtex4}

\usepackage{graphicx}
\usepackage{dcolumn}
\usepackage{bm}

\begin{document}

\title{Complementary Pair Density Wave and $d$-wave Checkerboard Order in High Temperature Superconductors}

\author{Kangjun Seo}
\affiliation{Department of Physics, Purdue University, West Lafayette, Indiana 47907, USA}

\author{Han-Dong Chen}
\affiliation{Department of Physics, University of Illinois at Urbana-Champaign, Urbana, IL 61801, USA}

\author{Jiangping Hu}
\affiliation{Department of Physics, Purdue University, West Lafayette, Indiana 47907, USA}

\date{\today}


\begin{abstract}
The competing orders in the particle-particle (P-P) channel and
the particle-hole (P-H) channel have been proposed separately to explain the
pseudogap physics in cuprates. By solving the Bogoliubov-deGennes
equation self-consistently, we show that there is a general
complementary connection between the $d$-wave checkerboard order
(DWCB) in the particle-hole (P-H) channel and the pair density wave order
(PDW) in the particle-particle (P-P) channel. A small pair density
localization generates DWCB  and PDW  orders simultaneously. The
result suggests that suppressing superconductivity locally or
globally through phase fluctuation should induce both orders in
underdoped cuprates. The presence of both DWCB and PDW  orders with
$4a \times 4a$ periodicity can explain the checkerboard modulation
observed in FT-STS from STM and the puzzling dichotomy between the
nodal and antinodal regions as well as the characteristic features
such as non-dispersive Fermi arc in the pseudogap state.
\end{abstract}

\pacs{74.25.Jb, 74.25.Dw, 74.72.-h}

\maketitle


\section{Introduction}
An important characteristic of strongly correlated electron systems
is the existence of different instabilities that
lead to  many competing orders. In high temperature superconductors,
besides the superconducting phase,  many competing orders, such as,
spin density wave (SDW)\cite{ZHANG1997, SACHDEV2004}, $d$-density wave
(DDW)\cite{NAYAK2000,CHAKRAVARTY2001}, pair density wave
(PDW)\cite{CHEN2002,CHEN2004}, stripe\cite{KIVELSON2003}, and so on,
have been proposed to explain various experimental observations.
Those competing orders can be generally classified into two
categories, the orders in particle-particle (P-P) channel and the orders in
the particle-hole (P-H) channel. So far, most of theoretical works in
cuprates  have focused on the effect of individual competing orders.
However,  the orders in two channels  are not completely independent
of each other.  In some cases, they  must be correlated.  In this
paper, we detailed study one of these examples, the extended $s$-wave PDW order in the
P-P channel and the $d$-wave checkerboard density
(DWCB) order in the P-H channel.

The motivation of this study  mainly comes from  the recent
experiments of Scanning Tunneling Microscopy(STM) in cuprates. These
experiments have revealed surprising yet important electronic
structures in the high temperature superconductors. The Fourier
transform scanning tunneling spectroscopies (FT-STS) from STM  have
captured two different general features in both momentum and energy
spaces~\cite{HOFFMAN2002,HOWALD2003,HANAGURI2004,MCELROY2003,MCELROY2003A,VERSHININ2004,MCELROY2005A,MCELROY2005,
FANG2006}. One feature is dispersive peaks in
FT-STS~\cite{MCELROY2003, MCELROY2003A}, interpreted as interference
patterns caused by elastic scattering of quasiparticles from
impurities~\cite{WANG2003}. The other is non-dispersive peaks,  a
checkerboard modulation observed in various different materials and
circumstances.
 The checkerboard structure was first discovered locally in BSCCO near a vortex core~\cite{PAN2000,HOFFMAN2002}.
  Then, it was found to be a characteristic of the large gap regions
  where the STM spectrum resembles that in the pseudogap phase~\cite{HOWALD2003,MCELROY2003A,MCELROY2005A}.
  Later, in the pseudogap phase, a similar checkerboard pattern was also observed~\cite{VERSHININ2004}.
  Finally, the STM studies of Ca$_{2-x}$Na$_x$CuO$_2$Cl$_2$ revealed the presence of a global
  long range commensurate checkerboard order independent of doping~\cite{HANAGURI2004}.

There have been various theoretical proposals to explain   the
non-dispersive checkerboard modulations. Most of these proposals are
related to the competing orders.  In these theories, the origin of
the  non-dispersive modulations are tied to the existence of
particular order parameters. The theories including pair density
modulation~\cite{CHEN2002,CHEN2004,CHEN2004A,TESANOVIC2004,TESANOVIC2005},
current density modulation~\cite{BENA2004,GHOSAL2005}, spin
modulation~\cite{SACHDEV2004} , stripe charge
modulation~\cite{KIVELSON2003,ROBERTSON2006}, and impurity
scattering~\cite{PODOLSKY2003} and so on.

 Among the proposed
mechanisms, the pair density wave (PDW) has been shown to capture
important characteristics of the checkerboard density modulation.
The mechanism of PDW derives from high pairing energy scale in
cuprates. It suggests that, unlike the superconductivity of normal
BCS type superconductors that can be destroyed by breaking Cooper
pairs, the superconductivity in cuprates can be more easily weakened
or destroyed by phase fluctuations than by pair breaking. Based on
this argument,  pair density localization~\cite{CHEN2002} was first
proposed to explain the local checkerboard modulation in the
presence of impurity or vortex. Later, a global pair density
wave(PDW) was proposed to explain the checkerboard physics in the
pseudogap state~\cite{VERSHININ2004,CHEN2004}. It has also been
shown that the symmetry of the tunneling intensity can distinguish
the pair density modulation from the conventional density
modulation~\cite{CHEN2004}. While the pair density modulation
provides a good understanding of the experimental results, the
theory does not cover two important characterizations of the tunneling experiments, namely, the dichotomy between nodal and antinodal regions and the Fermic arc.

Recently, we have proposed  a $d$-wave checkerboard density (DWCB)
order in P-H channel~\cite{SEO2007}.  The DWCB can be
viewed as a natural extension of the $d$-density wave(DDW) order
proposed to explain pseudogap
physics~\cite{NAYAK2000,CHAKRAVARTY2001}, and is only different from
the latter in terms of order wavevectors. We have shown that the
DWCB order must exist when the PDW order is present in the global
$d$-wave superconducting state. Moreover, we have demonstrated that
the DWCB  captures many general features  of the STM experimental
results.  It has been demonstrated that the DWCB order has little
effect on the density of state at low energy in the superconducting
phase, but has a strong effect on the STM spectrum around  the
superconducting gap at high energy. This result naturally explains
the puzzling dichotomy between the nodal and antinodal regions
observed in STM~\cite{MCELROY2005A} and angle resolved photoemission
spectroscopy (APRES)~\cite{ZHOU2004}. The DWCB order also preserves
in FT-STS spectrum at the wavevectors, $\mathbf{Q}=\{(\pi/2a,0),(0,
\pi/2a)\} $, the same symmetry as that observed in the experiments.
Moreover, the DWCB preserves the nodes in the single particle
spectrum, and generates a Fermi arc with little dispersion around
the nodal points at high temperature, which are consistent with the
results from ARPES. The Fermi arc has been a signature of the
pseudogap region, and has been proposed to explain the checkerboard
pattern observed in the pseudogap state~\cite{Chatterjee2006}. Thus,
the DWCB provides a physical origin of the Fermi arc.

In this paper, by solving the Bogoliubov-deGennes equation
self-consistently, we further demonstrate the results drawn in
ref.~\cite{SEO2007}.  We design a microscopic model with the space
modulated density-density interaction.   We show that  from the
self-consistent solutions of the Bogoliubov-deGennes equations, a
weak spatially modulated density-density interactions can generate the DWCB
and the PDW orders simultaneously in general. The combination  of these
two orders captures many important features of the STM experimental
results. We organized the rest of this paper as follows: In Sec.II
We introduce the $d$-wave checkerboard density (DWCB) order to study
the electronic states in the disordered $d$-wave superconducting
states, and the pseudogap phase at a temperature above $T_c$. In Sec.III we show how the complementary connection
between the density orderings in the
 particle-particle and the particle-hole channels are closely related in the order wavevector
 and the symmetries. In Sec.IV we study the orders in cuprates by calculating a full
 self-consistent BdG equation including DWCB and PDW in the $d$-wave superconducting state.
 In Sec.V we calculate the local density of states and the spectral weights in the
 presence of both DWCB and PDW and compare the recent experimental data.


\section{$d$-wave checkerboard density order in cuprates }

In this secion, we shall introduce the DWCB order and discuss its roles in different regions of the phase diagram in high temperature superconductors.
In ref.~\cite{SEO2007}, we have proposed a DWCB to explain the experimental results observed in the STM measurements in a global $d$-wave superconducting state. It has been shown that the presence of DWCB and competition with DSC can capture the physics in a disordered superconducting state and in the pseudogap phase at a temperature above $T_c$. Here we present a more detailed discussion and analysis. Firstly, we will show the average density of states(DOS) and the Fourier component at the order wavevector $\mathbf{Q} = \left\{ (\pi/2,0),(0,\pi/2)\right\}$ calculated in a mixed state of DSC and DWCB orders show good agreements with experiments such as the spatially modulated LDOS in the disordered superconductors. Secondly, we will show the presence of the DWCB order with the $d$-wave superconducting order at a higher temperature above $T_c$ plays an important role in understanding the physics of the pseudogap phase: an emergence of the energy-independent Fermi arcs above $T_c$ and the linear dependence of its length on temperature.

The mean-field Hamiltonian for the system where the DWCB coexists with $d$-wave superconducting order is given by
\begin{equation}
\mathcal{H}_{\text{MF}} = \mathcal{H}_0 + \mathcal{H}_{\text{DWCB}},
\label{h_dwcb}
\end{equation}
where
\begin{eqnarray}
\mathcal{H}_0 &=& \sum_{\mathbf{k},\sigma} \xi_\mathbf{k} c_{\mathbf{k}\sigma}^\dagger c_{\mathbf{k}\sigma}
        + \sum_\mathbf{k} \Delta_\mathbf{k} c_{\mathbf{k}\uparrow}^\dagger c_{-\mathbf{k}\downarrow}^                       \dagger + \text{h.c.} \\
\mathcal{H}_{\text{DWCB}} &=& \sum_{\mathbf{k},\mathbf{q},\sigma} W_{\mathbf{k},\mathbf{q}} c_{\mathbf{k}+\mathbf{q}\sigma}^\dagger c_      {\mathbf{k}\sigma} + \text{h.c.}
\end{eqnarray}
$\mathcal{H}_0$ is the mean-field Hamiltonian for the $d$-wave superconducting state and thus $\Delta_\mathbf{k} = \Delta_0/2 \left(\cos k_y - \cos k_y \right)$.
The DWCB has particular order wavevectors, $\mathbf{Q} = \left\{ (\pi/2,0),(0,\pi/2)\right\}$ and can be written as $W_{\mathbf{k},\mathbf{q}} = W_\mathbf{k} \delta_{\mathbf{q},\mathbf{Q}}$ with $W_\mathbf{k} = W_0 /2 \left( \cos k_x - \cos k_y \right)$.

\begin{figure}
\includegraphics[width=8cm]{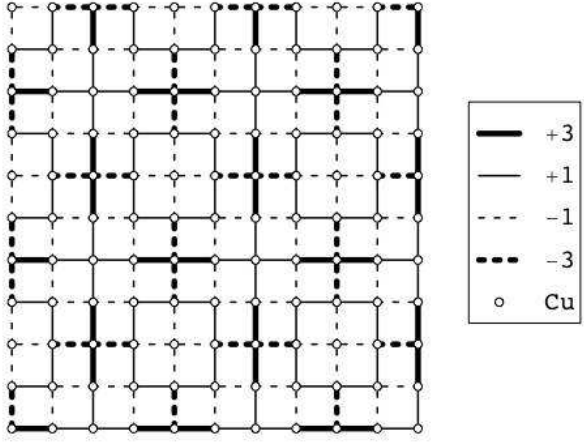}
\caption{
\label{fig_dwcb_real}
The configuration of the bond density of the DWCB order in the real space.
It is manifestly shown that the pattern has $4a \times 4a$
periodicity and $d_{x^2-y^2}$ symmetry.
}
\end{figure}

 To obtain a clear picture of the DWCB, we rewrite $\mathcal{H}_{\text{DWCB}}$ in the real space configuration:

\begin{eqnarray}
 &&\mathcal{H}_{\text{DWCB}} = \sum_{\mathbf{r}}
\text{Im}(W_0)\left[\left(\cos \frac{\pi x}{2}+ \sin\frac{\pi
 x}{2}\right)\hat J_{\mathbf{r}}^x-x\leftrightarrow y
\right]\nonumber\\
&&+\text{Re}(W_0) \left[\left(\cos \frac{\pi x}{2}-\sin\frac{\pi
 x}{2}+2\cos \frac{\pi y}{2}\right)\hat B_{\mathbf{r}}^x-x\leftrightarrow y\right],
\label{h_eff_int}
\end{eqnarray}
 where $\mathbf{r}=(x,y)a$, $\hat B_{\mathbf{r}}^{x(y)} =\sum_\sigma (c^\dagger_{\mathbf{r},\sigma}c_{\mathbf{r}+a\hat
 x(\hat
 y),\sigma} +h.c.)$ is the density operator defined in the nearest neighbor bond between $\mathbf{r}$ and $\mathbf{r}+a\hat x(\hat y)$ with $\hat x (\hat y)$
 unit vectors along $x(y)$ directions, and $\hat J_{\mathbf{r}}^{x(y)} =i\sum_\sigma (c^\dagger_{\mathbf{r},\sigma}c_{\mathbf{r}+a\hat
 x(\hat
 y),\sigma} -h.c.)$ is the
 current density operator defined in the same bond as $\hat B_{\mathbf{r}}^{x(y)}$.
 Fig.~\ref{fig_dwcb_real} shows a static pattern of the bond strength of the DWCB order, $\langle \hat{B}_{\mathbf{r}} \rangle$.
 It is clear that the DWCB order defined in Eq.(\ref{h_eff_int}) has $4a\times 4a$ periodicity and $d_{x^2-y^2}$ symmetry.
Similar order parameters have been mentioned in Ref.\cite{PODOLSKY2003}.

\subsection{ DWCB in the disordered Superconducting state }

Based on the mean-field Hamiltonian, Eq.(\ref{h_dwcb}), we have calculated the averaged local density of states, $\rho(\omega)$, and the Fourier component  of local density states(LDOS) at $\mathbf{Q} = \{
(\pi/2,0),(0,\pi/2)\}$, $\rho_\mathbf{Q} (\omega)$, in two band dispersions. The calculation results turned out to be rather general and
insensitive to the bare band structures.
The effective mean-field Hamiltonian, Eq.(\ref{h_dwcb}), can be rewritten by using Nambu formalism:
\begin{equation}
\mathcal{H}_{\text{MF}} = \sum_\mathbf{k} \psi_\mathbf{k}^\dagger H(\mathbf{k}) \psi_\mathbf{k}
\end{equation}
where $\psi_\mathbf{k} = \left( c_{\mathbf{k}\uparrow}, c_{\mathbf{k}+\mathbf{Q}\uparrow}, c_{-\mathbf{k}\downarrow}^\dagger, c_{-\mathbf{k}-\mathbf{Q}\downarrow}^\dagger \right)^\dagger$, and
\begin{equation}
H(\mathbf{k}) = \left( \begin{array}{cccc}
\xi_\mathbf{k} & W_\mathbf{k} & \Delta_\mathbf{k} & 0 \\
W_\mathbf{k}^\ast & \xi_{\mathbf{k}+\mathbf{Q}} & 0 & \Delta_{\mathbf{k}+\mathbf{Q}} \\
\Delta_\mathbf{k}^\ast & 0 & -\xi_{-\mathbf{k}} & W_{\mathbf{k}+\mathbf{Q}}^\ast \\
0 & \Delta^{\ast}_{\mathbf{k}+\mathbf{Q}} & W_{\mathbf{k}+\mathbf{Q}} & -\xi_{-\mathbf{k} + \mathbf{Q}}
\end{array}
\right)
\end{equation}
Then the retarded Green function is given by
\begin{equation}
\mathbf{G}^{-1}(\mathbf{k},\omega) = \left( \omega + i\eta \right) \mathbf{I} - H(\mathbf{k}), 
\end{equation}
where $\mathbf{I}$ is the identity matrix with the same rank with $H(\mathbf{k} )$.
The averaged density of states(DOS) and the Fourier component at $\mathbf{Q}$ can be calculated as the following, respectively:
\begin{eqnarray}
\rho (\omega ) &=& - \frac{1}{\pi}\sum_\mathbf{k} \text{Im} \mathbf{G}_{11}(\mathbf{k},\omega) \\
\rho_\mathbf{Q} (\omega) &=& - \frac{1}{\pi} \sum_\mathbf{k} \text{Im} \mathbf{G}_{12}(\mathbf{k},\omega)
\end{eqnarray}

Firstly, we performed
calculations in the particle-hole symmetric case, where the band
dispersion is given by
\begin{equation}
\xi_\mathbf{k} = -t/2 \left( \cos k_x + \cos k_y \right) - \mu
\label{simple_dispersion}
\end{equation}
We chose $t = 125$meV and $\mu = 0$. $\Delta_{0}
= 40$meV, which is relevant for underdoped BSCCO. The imaginary part
of the self energy $\eta = 5$meV was used for the numerical
calculation.

\begin{figure}
\includegraphics[width=8cm]{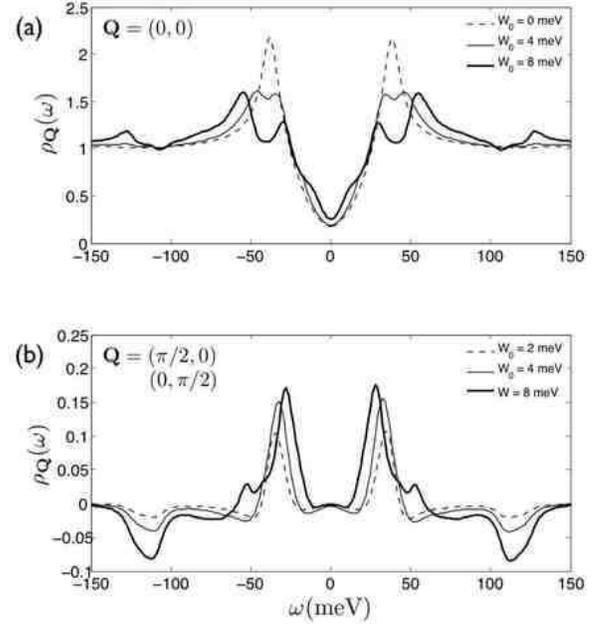}
\caption{
\label{fig_dos_dosq_simple}
(a) Averaged local density of states (LDOS) are
plotted for various DWCB orders: $W_0 = 0, 4$, and 8meV with $t=-125$meV, $t' = \mu = 0$.
(b) The Fourier components of LDOS at $\mathbf{Q} = \left\{(\pi/2,0),(0,\pi/2)\right\}$.
}
\end{figure}

Fig.~\ref{fig_dos_dosq_simple}(a) shows the averaged DOS
normalized by one of the non-interacting Fermi liquids. In the absence of
DWCB order, i.e. $W_0 = 0$, there are  sharp coherence peaks at the
energy of superconducting gap, as expected. As DWCB order develops, the coherence
peaks located at 40meV are suppressed, while the spectrum at low
energy remains unchanged. From $W_0 = 4 \text{meV}$ the prominent peak begins to appear within the superconducting gap. Note that even small DWCB made a strong effect on the spectrum at high energy as $W_0$ increases.

Fig.~\ref{fig_dos_dosq_simple}(b) shows the Fourier
components of LDOS at the wavevectors, $\mathbf{Q} = \left\{(\pi/2,0), (0,\pi/2)\right\}$. It is known\cite{CHEN2004} that $\rho _\mathbf{Q} (\omega ) = \rho _\mathbf{Q} (-\omega)$ for a bond-centered P-H pairing, while $\rho _\mathbf{Q} ( \omega) = - \rho _\mathbf{Q} ( -\omega )$ for a site-centered P-H pairing such as a conventional charge density wave(CDW).
As expected from the fact that the DWCB order is bond-centered, $\rho _\mathbf{Q} ( \omega)$ is even with respect to $\omega$, and it shows good agreement with experiment~\cite{HOWALD2003}.

\begin{figure}
\includegraphics[width=8cm]{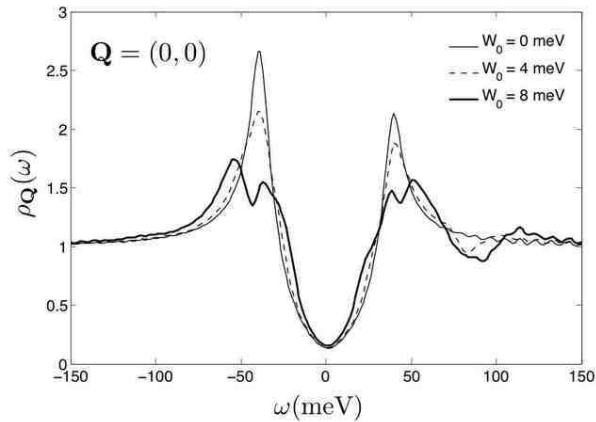}
\caption{
\label{fig_dos_norman}
Averaged local density of states with the finite chemical
potential included in the band dispersion provided by Norman {\it et
al}~\cite{NORMAN1995}. }
\end{figure}

We repeat our
calculations with the more realistic band dispersion provided by Norman {\it et
al.}~\cite{NORMAN1995} and the result is displayed at
Fig.~\ref{fig_dos_norman}. The band energy dispersion is now modified as such
\begin{eqnarray}
\xi_\mathbf{k} &=& t_1/2 (\cos k_x + \cos k_y ) + t_2 \cos k_x \cos k_y \nonumber\\
     &+& t_3/2 (\cos 2 k_x + \cos 2 k_y )\nonumber \\
     &+& t_4/2 ( \cos 2 k_x \cos k_y + \cos k_x \cos 2 k_y )\nonumber \\
     &+& t_5 \cos 2k_x \cos 2 k_y -\mu,
\end{eqnarray}
 where $t_1 = -0.5951$eV, $t_2 = 0.1636$eV, $t_3 = -0.0519$eV,
$t_4 = -01117$eV, and $t_5 = 0.0510$eV~\cite{NORMAN1995}. The chemical potential $\mu$
is now set to -0.1660eV. Compared with the P-H symmetric
case, the effect of DWCB on the averaged DOS and the Fourier component of LDOS at $\mathbf{Q}$ is insensitive to the energy
band structure. Qualitatively the numerical results are strikingly
consistent with experimental results~\cite{FANG2006}, and
the large gap region can be interpreted in terms of the coexistence of
weak(8-12meV) DWCB and DSC orders.

\subsection{DWCB in the Pseudogap state}

The results of LDOS have demonstrated the consistency between the coexistence of the  DWCB and DSC orders
and the STM experimental results in the disordered superconducting state. Now we will show that the presence of the DWCB order also captures important physics in the
pseudogap phase at a temperature above $T_c$. 
While there still have been hot debates over the interpretation of the origin of the the pseudogap phase, we will show how many experimental observations in this phase can be explained by interpreting the pseudogap phase as the mixed state of the DWCB and DSC orders. 
In this paper, we will focus on two features of the Fermi arcs developed from the nodal point along the Fermi surface: non-dispersive energy-independence and linear temperature dependence.

The autocorrelation of ARPES data from BSCCO show non-dispersive peaks in momentum space arising from the tips of the Fermi arcs in the pseudogap phase, while the superconducting state shows dispersion with binding energy~\cite{Chatterjee2006}.
For the temperature dependence of the Fermi arcs in the pseudogap phase, it is known that its length increases linearly with temperature in the range between $T_c$ and $T^{\ast}$, below which the material is believed to be in  the pseudogap phase~\cite{Kanigel2006}. 
If the pseudogap phase is strongly
connected to phase fluctuations of $d$-wave superconductivity, the
single particle spectrum should reflect the DWCB order. Therefore, a
robust Fermi arc feature should exist in the mixed DWCB and DSC
phases at high temperature.
We will show that this is indeed the case.

\begin{figure}
\includegraphics[width=8cm]{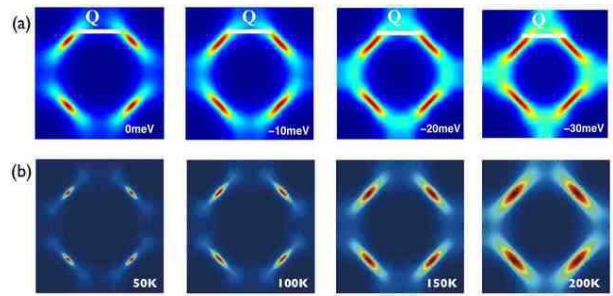}
\caption{
\label{fig_Akw_as_E_T}
The spectral function given by Eq.(\ref{spectral_weight}) based on the energy dispersion, (\ref{simple_dispersion}).
(a) The energy dependence of the Fermi arc along the Fermi surface.
The white bars with the magnitude of $|\mathbf{Q}|=\pi/2$ are displayed to show the Fermi arc is non-dispersive.
The deviations for the energies ($\omega = 0, -10, -20, -30$ meV) from $|\mathbf{Q}|$ are negligibly small by 8\%, 2\%, 0\%, and 13\% of $|\mathbf{Q}|$, respectively.
(b) In the prsence of DSC($\Delta_{0} = 40$ meV) and DWCB($W_0=8$meV),   the spectral weights, $A (\mathbf{k}, \omega=0meV)$, are plotted with varying temperature
($T = 50K, 100K, 150K$, and $200K$).
}
\end{figure}

In order to illustrate the emergence of the non-dispersive Fermi arc in the pseudogap state with DWCB order coexistent with DSC, we have calculated the spectral function $A(\mathbf{k},\omega)$ as given by the imaginary part of the retarded Green function:
\begin{equation}
A(\mathbf{k},\omega) = -\frac{1}{\pi} \text{Im} \mathbf{G}_{11}(\mathbf{k},\omega)
\label{spectral_weight}
\end{equation}
First we have studied the energy dependence of the Fermi arcs in the pseudogap state.
 We have plotted the spectral function $A(\mathbf{k},\omega)$ in the first Brillouin zone in Fig.~\ref{fig_Akw_as_E_T} based on the energy dispersion, Eq.(\ref{simple_dispersion}).
In Fig.~\ref{fig_Akw_as_E_T}(a), the Fermi arcs are in red, and one of the $d$-wave checkerboard wavevectors, $\mathbf{Q}=(\pi/2,0)$, is shown as a white bar.
The scattering wavevectors connecting the tips of each arc are nearly equal to the order wavevectors of DWCB, $|\mathbf{Q}| = \pi/2$, which is consistent with the non-dispersive Fermi arc in experimental observations.
The small dispersion is negligible when compared with the elongation of the gapless part along the Fermi surface from the nodal point in the DSC state without DWCB order.
It may depend on the band width of the calculation. When including more hopping terms in the energy band, the dispersion shown in the simple band calculation will be reduced.

Now let us consider the
temperature dependence of $A(\mathbf{k},\omega)$ at the Fermi level($\omega = 0$meV).
It can be calculated in the pseudogap phase by taking the temperature dependence as an effect from the self energy, $\eta$.
 We have plotted the spectral function $A(\mathbf{k},\omega)$ as a function of the temperature in the first Brillouin zone in Fig.~\ref{fig_Akw_as_E_T}(b).
 At very low
temperature the Fermi surface is gapped except at the nodal point,
$(\pi/2,\pi/2)$. As the temperature rises, the nodal points grow
significantly along the Fermi surface with slight broadening
in the direction perpendicular to the Fermi surface. We will show the detailed calculations and discuss more in section V.

%


\section{ Complementary connection between the orders in the P-P and P-H channels }

In this section, we will give a general argument regarding the complementary connection between the orders in the particle-particle and the particle-hole channels in the disordered $d$-wave superconducting state at zero temperature as well as in the pseudogap phase at high temperature above $T_c$. 
As shown above,  a $d$-wave checkerboard order (DWCB) in the P-H channel, $\langle c_{\mathbf{k}\sigma}^\dagger c_{\mathbf{k}+\mathbf{Q}\sigma} \rangle = \Phi f(\mathbf{k}) $ with $\mathbf{Q} =\{ (\pi/2,0),(0, \pi/2)\}$ and $f(\mathbf{k}) = \cos k_x - \cos k_y$, can explain the experiments on cuprates in the disordered DSC state and in the pseudogap phase.  Preserving the same symmetry in FT-STS spectrum as that observed in experiments, the DWCB generates Fermi arcs with little dispersion with the binding energy around nodal points at high temperature above $T_c$.  Since the Fermi arc has been a signature of the pseudogap, the DWCB provides a direct link between the competing order and the pseudogap physics.

The DWCB, however, is not a completely independent mechanism. In fact, there are some intimate connections with other orders in the P-P channel. Due to the existence of a complementary connection between orders in both channels, the DWCB is directly connected to the PDW order. The angular symmetry of one orderings is determined by the combination of the angular symmetries of DSC and the complementary order. In this section, we will study how these density orderings in both channels are related in the modulation wavevector and the symmetry.

\subsection{ The Connection between PDW and DDW }

 To show the connection, we will use the example of the $d$-density wave (DDW) order. Since the density order in the P-H channel has been well studied, examining the complementary connected ordering in the P-P channel will be a preliminary step to generalize the complementary connection in the case of DWCB.

The mean field Hamiltonian in the DSC state coexisting with DDW can be written as
\begin{eqnarray}
\mathcal{H} &=& \sum_{\mathbf{k}\sigma} \xi_\mathbf{k} c_{\mathbf{k}\sigma}^\dagger c_{\mathbf{k}\sigma} +  iW_\mathbf{k} c_{\mathbf{k}\sigma}^\dagger c_{\mathbf{k}+\mathbf{Q}\sigma} + \sum_\mathbf{k} \Delta_\mathbf{k} c_{\mathbf{k}\uparrow}^\dagger c_{-\mathbf{k}\downarrow}^\dagger + \text{h.c.} \nonumber\\
&=&  \sum_\mathbf{k} \psi^\dagger (\mathbf{k}) A(\mathbf{k}) \psi (\mathbf{k}),
\end{eqnarray}
where $\psi^\dagger (\mathbf{k}) = ( \begin{array}{llll} c_{\mathbf{k}\uparrow} & c_{\mathbf{k}+\mathbf{Q}\uparrow} &                                                                            c^\dagger_{-\mathbf{k}\downarrow} & c^\dagger_{-\mathbf{k}-\mathbf{Q}\downarrow} \end{array} )^\dagger$,
and
\begin{equation}
    A(\mathbf{k}) = \left( \begin{array}{cccc}
        \xi_\mathbf{k} & i W_\mathbf{k} & \Delta_\mathbf{k} &  0 \\
        -i W_\mathbf{k} & -\xi_\mathbf{k} & 0 & -\Delta _\mathbf{k} \\
        \Delta^\ast _\mathbf{k} & 0 & -\xi _\mathbf{k} & i W _\mathbf{k} \\
        0 & -\Delta^\ast _\mathbf{k} & -i W _\mathbf{k} & \xi _\mathbf{k}\end{array}\right),
\end{equation}
where $c_{\mathbf{k}\sigma}^\dagger$ and
$c_{\mathbf{k}\sigma}$ are the creation operator and destruction operator of
the single particle with spin $\sigma$, respectively. $\xi _\mathbf{k} $ is the energy
dispersion of the single particle. $\Delta _\mathbf{k}= \Delta_{0} /2
\left( \cos k_x - \cos k_y \right)$ and $W_\mathbf{k} = W_0/2 \left(\cos k_x - \cos k_y \right)$ are the $d$-wave supeprconducting order (DSC) and the DDW, respectively.
Then the Hamiltonian is diagonalized by the following transformation:

\begin{widetext}
\begin{equation}
    \left( \begin{array}{l} c_{\mathbf{k}\uparrow} \\ c_{\mathbf{k}+\mathbf{Q}\uparrow} \\                                                                              c^\dagger_{-\mathbf{k}\downarrow} \\ c^\dagger_{-\mathbf{k}-\mathbf{Q}\downarrow} \end{array}\right) = \left( \begin{array}{cccc}
               \frac{-iW_\mathbf{k}}{\sqrt{2E_\mathbf{k} ( E_\mathbf{k} - \xi _\mathbf{k} )}} & \frac{\xi_\mathbf{k} +E_\mathbf{k}}{\sqrt{2E_\mathbf{k} ( E_\mathbf{k} + \xi _\mathbf{k} )}} &
               \frac{-i W_\mathbf{k}}{\sqrt{2E_\mathbf{k} ( E_\mathbf{k} + \xi _\mathbf{k} )}} & \frac{\xi_\mathbf{k}-E_\mathbf{k}}{\sqrt{2E_\mathbf{k} ( E_\mathbf{k} - \xi _\mathbf{k} )}} \\
               \frac{\xi_\mathbf{k}-E_\mathbf{k}}{\sqrt{2E_\mathbf{k} ( E_\mathbf{k} - \xi _\mathbf{k} )}} & \frac{-i W_\mathbf{k}}{\sqrt{2E_\mathbf{k} ( E_\mathbf{k} + \xi _\mathbf{k} )}}
               &  \frac{\xi_\mathbf{k} +E_\mathbf{k}}{\sqrt{2E_\mathbf{k} ( E_\mathbf{k} + \xi _\mathbf{k} )}}  &  \frac{-iW_\mathbf{k}}{\sqrt{2E_\mathbf{k} ( E_\mathbf{k} - \xi _\mathbf{k} )}} \\
               0 & \frac{\Delta _\mathbf{k}}{\sqrt{2E_\mathbf{k} ( E_\mathbf{k} + \xi _\mathbf{k} )}} & 0 & \frac{\Delta _\mathbf{k}}{\sqrt{2E_\mathbf{k} ( E_\mathbf{k} - \xi _\mathbf{k} )}} \\
               \frac{\Delta _\mathbf{k}}{\sqrt{2E_\mathbf{k} ( E_\mathbf{k} - \xi _\mathbf{k} )}} & 0 & \frac{\Delta _\mathbf{k}}{\sqrt{2E_\mathbf{k} ( E_\mathbf{k} + \xi _\mathbf{k} )}} &  0
               \end{array}
               \right)
               \left( \begin{array}{l} \gamma_1 (\mathbf{k}) \\ \gamma_2 (\mathbf{k}) \\ \gamma_3 (\mathbf{k}) \\ \gamma_4 (\mathbf{k}) \end{array} \right)
\end{equation}
\end{widetext}
where $E _\mathbf{k} = \sqrt{ \xi_\mathbf{k}^2 + |\Delta_\mathbf{k}|^2 +  W_\mathbf{k}^2 }$.
The eigenvalues $E_\alpha (\mathbf{k})$ corresponding to the eigenvectors $\gamma_\alpha (\mathbf{k})$ are given by
\begin{eqnarray}
&& E_1 (\mathbf{k}) = E_2 (\mathbf{k}) = + E _\mathbf{k} \\
&& E_3 (\mathbf{k}) = E_4 (\mathbf{k}) = - E _\mathbf{k}.
\end{eqnarray}
 The ground state $|\Omega\rangle$ is defined by the following conditions:
\begin{eqnarray}
&& \gamma_1 (\mathbf{k}) |\Omega\rangle = \gamma_2 (\mathbf{k}) |\Omega\rangle = 0, \\
&& \gamma_3^\dagger (\mathbf{k}) |\Omega\rangle = \gamma_4^\dagger (\mathbf{k}) |\Omega\rangle = 0.
\end{eqnarray}
Then non-zero expectation values of the bilinear operators of $c$'s are
\begin{eqnarray}
\langle c_{\mathbf{k}\uparrow}^\dagger c_{\mathbf{k}\uparrow} \rangle &=& v_\mathbf{k}^2 \frac{|\Delta_\mathbf{k}|^2 }{ |\Delta_\mathbf{k}|^2 + W_\mathbf{k}^2 } \\
\langle c_{\mathbf{k}\uparrow}^\dagger c_{-\mathbf{k}\downarrow}^\dagger \rangle &=& \frac{\Delta_\mathbf{k}}{2E_\mathbf{k}} \\
\langle c_{\mathbf{k}\uparrow}^\dagger c_{\mathbf{k}+\mathbf{Q}\uparrow} \rangle &=& \frac{ i W_\mathbf{k}}{E_\mathbf{k}} \\
 \langle c_{\mathbf{k}\uparrow}^\dagger c_{-\mathbf{k}-\mathbf{Q}\downarrow}^\dagger \rangle &=& v_\mathbf{k}^2 \frac{iW_\mathbf{k} \Delta_\mathbf{k} }{ |\Delta_\mathbf{k}|^2 + W_\mathbf{k}^2 }
 \end{eqnarray}
 where $v_\mathbf{k}^2 = \frac{1}{2} \left( 1 - \frac{\xi_\mathbf{k}}{E_\mathbf{k}} \right)$.
Note that in a DSC state without DDW, or $W_\mathbf{k} = 0$,
\begin{equation}
\langle c_{\mathbf{k}\uparrow}^\dagger c_{\mathbf{k}\uparrow} \rangle = \frac{1}{2} \left( 1 - \frac{\xi_\mathbf{k}}{\sqrt{\xi_\mathbf{k}^2+|\Delta_\mathbf{k}|^2}} \right)
\end{equation}

Therefore, in the mixed state of the DSC, $\langle c_{\mathbf{k}\uparrow}^\dagger c_{-\mathbf{k}\downarrow}^\dagger \rangle \propto \Delta_\mathbf{k} $,  and DDW, $\langle c_{\mathbf{k}\sigma}^\dagger c_{\mathbf{k}+\mathbf{Q}\sigma}\rangle \propto i W_\mathbf{k}$, a PDW order with the same wavevector $\mathbf{Q}$ is expected to exist naturally, whose symmetry is an extended $s$-wave manifestly given by
\begin{equation}
\langle c_{\mathbf{k}\uparrow}^\dagger c_{-\mathbf{k}-\mathbf{Q}\downarrow}^\dagger\rangle \propto i W_\mathbf{k} \Delta_\mathbf{k} = \frac{iW_0 \Delta_0 }{4} \left( \cos k_x - \cos k_y \right)^2.
\end{equation}
 This indicates that the mixed state of DDW and DSC leads to the existence of PDW. This argument can also be applied to the state of DSC coexisting with PDW leading to DDW. Therefore the mixed state of DSC and DDW is nothing but a complementarily connected description of the mixed state of DSC and PDW.

\subsection{ The Connection  between PDW and DWCB}

Given the close resemblance between DDW and DWCB, it is natural to expect that similar results also hold for the DWCB order, since the difference between DDW and DWCB is the order wavevector $\mathbf{Q}$. Due to the high energy scale of superconducting gap in HTS, SC can be destroyed or suppressed by phase fluctuations. Carrying out a similar analysis as Gossamer superconductors~\cite{Laughlin2002}, we can show how the complementary connection between PDW and DWCB arises in the low energy
effective Hamiltonian that describes a system with pair modulation
induced by phase fluctuations.

 We start
from a BCS Hamiltonian on a 2D square lattice,

\begin{equation}
\mathcal{H}_{\text{BCS}} = \sum_{\mathbf{k}, \sigma}\xi_\mathbf{k} c_{\mathbf{k},\sigma}^\dagger
c_{\mathbf{k},\sigma} + \sum_{\mathbf{k}} \Delta_{\mathbf{k}} c_{\mathbf{k},\uparrow}^\dagger
c^\dag_{-\mathbf{k},\downarrow} +h.c.\label{h_bcs},
\end{equation}
The superconducting vacuum is constructed by the Cooper pairs with opposite momentum and spin:

\begin{equation}
|\Psi_{\text{BCS}}\rangle=\hat U_{\text{BCS}}|0\rangle \propto e^{\sum_\mathbf{k}\alpha_\mathbf{k} c_{\mathbf{k},\uparrow}^\dagger c^\dag_{-\mathbf{k},\downarrow}} |0\rangle,
\end{equation}
where $|\alpha_\mathbf{k}|^2=\frac{E_\mathbf{k}-\xi_\mathbf{k}+\mu}{E_\mathbf{k} + \xi_\mathbf{k}-\mu}$ and $E_\mathbf{k} = \sqrt{\xi_\mathbf{k}^2+|\Delta_\mathbf{k}|^2}$.
A local pair fluctuation induced by disorder such as a vortex, impurities, or other factors will lead to a new ground state with nonzero supercurrent due to the uncertainty principle. In other words, a new ground state $|\Psi\rangle$ can be obtained from the BCS ground state $|\Psi_{\text{BCS}}\rangle$ by applying a
boost of the total momentum $\mathbf{q}$ of the Cooper pairs,

\begin{equation}
|\Psi \rangle = e^{\eta \sum_{\mathbf{k},\mathbf{q}} ( \Phi_\mathbf{q} c_{\mathbf{k}\uparrow}^\dagger c_{-\mathbf{k}+\mathbf{q}\downarrow}^\dagger - \Phi_\mathbf{q}^\ast c_{-\mathbf{k}+\mathbf{q}\downarrow} c_{\mathbf{k}\uparrow}  ) } |\Psi_{\text{BCS}}\rangle  ,
\end{equation}
where $\eta$ is a small parameter and
$\Phi_\mathbf{q}$ is the structural factor of pair fluctuations determined phenomenologically. For a commensurate checkerboard modulation, $\Phi_\mathbf{q}$ will be peaked at the related total momenta $\mathbf{q}$ of the pair, that is $\mathbf{q} = \mathbf{Q}= \{( \pi/2,0),(0,\pi/2)\}$: $\Phi_\mathbf{q} \sim \delta_{\mathbf{q},\mathbf{Q}}$.

The effective Hamiltonian associated with $|\Psi \rangle$ as a ground state can be obtained as, up to a first order $\eta$,

\begin{equation}
  \mathcal{H}_{\text{eff}} =  \mathcal{H}_{\text{BCS}}+ \mathcal{H}_{\text{kin}} + \mathcal{H}_{\text{int}} + O(\eta^2),\label{eff_ham}
\end{equation}
where $\mathcal{H}_{\text{kin}}$ and $\mathcal{H}_{\text{int}}$ are additional terms generated by boosted kinetic and pair interaction terms, respectively, in Eq. (\ref{h_bcs}): for a commensurate checkerboard modulation,

\begin{eqnarray}\nonumber
    &\mathcal{H}_{\text{kin}} = -\eta \sum_{\mathbf{k},\mathbf{q}}  \xi_\mathbf{k} \Phi_\mathbf{q}
    (c^\dagger_{\mathbf{k},\uparrow}c^\dagger_{-\mathbf{k}+\mathbf{q},\downarrow}  +
    c^\dagger_{-\mathbf{k}+\mathbf{q},\uparrow}c^\dagger_{\mathbf{k},\downarrow})+h.c.\\    &= \sum_\mathbf{k} \Delta_{\mathbf{k},\mathbf{Q}}(c^\dagger_{\mathbf{k},\uparrow}c^\dagger_{-\mathbf{k}+\mathbf{q},\downarrow}  +
    c^\dagger_{-\mathbf{k}+\mathbf{q},\uparrow}c^\dagger_{\mathbf{k},\downarrow})+h.c. ,
\end{eqnarray}
where $\Delta_{\mathbf{k},\mathbf{Q}} = -\eta \sum_q \xi_\mathbf{k} \Phi_q \sim \xi_\mathbf{k} \delta_{\mathbf{q},\mathbf{Q}}$ and it represents an extended $s$-wave density order in the P-P channel, or PDW. And

\begin{eqnarray}\nonumber
  \mathcal{H}_{\text{int}} &=& \eta \sum_{\mathbf{k},\mathbf{q},\sigma} \Phi_\mathbf{q} \Delta_\mathbf{k}
  c_{\mathbf{k},\sigma}^\dagger c_{\mathbf{k}+\mathbf{q},\sigma}   + h.c.\\
  &=& \sum_{\mathbf{k},\sigma} W_\mathbf{k} c_{\mathbf{k},\sigma}^\dagger c_{\mathbf{k}+\mathbf{q},\sigma}   + h.c.,
  \end{eqnarray}
    where $W_\mathbf{k} = \eta \sum_\mathbf{q} \Phi_\mathbf{q} \Delta_\mathbf{k} \sim \Delta_\mathbf{k} \delta_{\mathbf{q},\mathbf{Q}}$ and it represents a $d$-wave density order in the P-H channel, DWCB.
The above derivation tells us that a $d$-wave-like ordering in the P-H channel
 effectively leads to Cooper pairs with a finite center of mass momentum.


\section{ Self-consistent calculation for pdw and dwcb }

Due to the complementary connection between PDW and DWCB, it is
suggestive to calculate a fully self-consistent Bogoliubov-de Gennes (BdG) equation
including both orders simultaneously. It has been argued that
long-range interaction between charge carriers is very important in
cuprates. PDW can be induced by the long range interaction between
hole-pairs. In this section, we will show that a solution of a
self-consistent calculation with modulating pair potential and
nearest neighbor interaction can lead to  orderings in both P-P and
P-H channels.

We start from a full model Hamiltonian on a 2D lattice,

\begin{eqnarray}
\mathcal{H} &=& -\frac{1}{2}\sum_{i,j,\sigma} \left[t_{ij}c^\dag_{i\sigma} c_{j\sigma}^{} + h.c.\right] + \sum_{i,j} V_{ij}n_{i} n_{j}\nonumber\\
&&-\mu\sum_in_i,
\end{eqnarray}
where $n_i = n_{i\uparrow} + n_{i\downarrow}$.

\begin{figure}
\includegraphics[width=8cm]{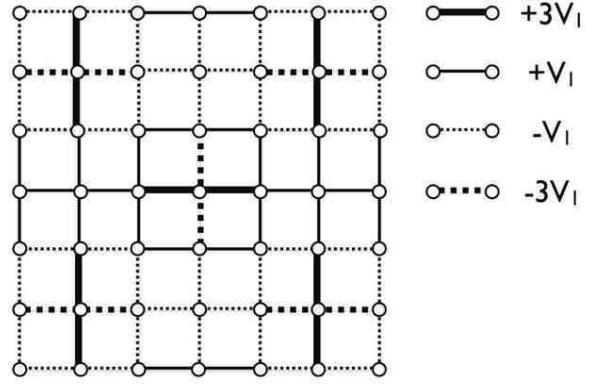}
\caption{
A modulating nearest-neighbor interaction, $\delta V_{ij}$, is plotted in a 2D square lattice.
\label{fig_vnn_modul}
}
\end{figure}

In particular, we will consider nearest-neighbor hopping, $t_{ij}= t$ and an attractive nearest-neighbor interaction, $V_{ij}$. The chemical potential $\mu$ is chosen such that it is P-H symmetric case. The nearest-neighbor interactions between two opposite spins are considered to be attractive and modulating. They are given by a negative constant pair potential and a long-range interaction modulated by $\mathbf{Q}=\{(\pi/2,0),(0,\pi/2)\}$,

\begin{eqnarray}\nonumber
V_{ij}  &=& -V_{0} + \delta V_{ij} \\
        &=& -V_0 + \sum_\mathbf{Q} \Delta V_{ij} ( \cos \mathbf{Q}\cdot \mathbf{r}_i + \cos
\mathbf{Q}\cdot \mathbf{r}_j),
\end{eqnarray}
where \begin{equation}
\Delta V_{ij} = \left\{ \begin{array}{ll}
                    +V_1, & j = i \pm \hat{x}\\
                    -V_1, & j = i \pm \hat{y}
                   \end{array}
           \right.
\end{equation}
with $V_0$ and $V_1$ positive constants.
While the local behavior of $\delta V_{ij}$ seems to be anisotropic at the center and $(\pm 2a,\pm 2a)$ as seen in Fig.~\ref{fig_vnn_modul}, the global behavior is isotropic in a sense that there is no preferential direction for attraction enhanced by negative modulation or weakened by positive modulation. The anisotropic modulation is rotationally invariant with the combination of translation by $4a $.


Starting with this Hamiltonian, let us derive the BdG equations by introducing mean-field decoupling of the nearest neighbor interaction terms.
The quartic term $n_{i\sigma} n_{j\sigma'}$ has three different types of mean-field decouplings
\begin{eqnarray}
&&c^\dag_{i\sigma} c^{}_{i\sigma} c^\dag_{j\sigma'} c^{}_{j\sigma'}
\nonumber\\\Rightarrow&&
\langle c^\dag_{i\sigma} c^{}_{i\sigma}\rangle c^\dag_{j\sigma'} c^{}_{j\sigma'}
+\langle c^\dag_{j\sigma'} c^{}_{j\sigma'}\rangle c^\dag_{i\sigma} c^{}_{i\sigma}\nonumber\\ &&
+\langle c^\dag_{i\sigma} c^{\dag}_{j\sigma'}\rangle c^{}_{j\sigma'} c^{}_{i\sigma}
+ c^\dag_{i\sigma} c^{\dag}_{j\sigma'} \langle c^{}_{j\sigma'} c^{}_{i\sigma}\rangle \nonumber\\ &&
-\langle c^\dag_{i\sigma} c^{}_{j\sigma'}\rangle c^{\dag}_{j\sigma'} c^{}_{i\sigma}
- c^\dag_{i\sigma} c^{}_{j\sigma'} \langle c^{\dag}_{j\sigma'} c^{}_{i\sigma}\rangle.
\end{eqnarray}
The first two terms are nothing but the chemical potential terms. The decouplings in the second line are the usual BCS decouplings. The last two terms are decouplings in the P-H channel, which are just the exchange terms. For interactions between two particles with same spins, these exchange terms effectively modified hopping terms. For interactions between two particles with opposite spins, it corresponds to a bond-centered spin density wave. From our full Hamiltonian, we can argue that the spin rotational symmetry is not broken. Thus the SDW decouplings can be safely ignored. Also, since we are interested in $d$-wave pairing, the BCS decouplings of triplet pairing will be dropped, too. We then arrive at the following MF Hamiltonian,
\begin{eqnarray}
\label{eqn:mfh}
\mathcal{H}_{MF}&=&-\frac{1}{2}\sum_{\langle i,j\rangle\sigma} \left[(t + W_{ij})c^\dag_{i\sigma} c_{j\sigma}^{} + h.c.\right]\nonumber\\
  &+& \sum_{\langle i,j\rangle} \left[\Delta^{(1)}_{ij}c^\dag_{i\uparrow}c^\dag_{j\downarrow}+h.c.\right]\\\nonumber
  &+& \sum_{\langle i,j\rangle} \left[\Delta^{(2)}_{ij}c^\dag_{i\uparrow}c^\dag_{j\downarrow}+h.c.\right]\\\nonumber
&-&\sum_{\langle i,j \rangle \sigma} \mu_{ij\sigma} n_{i\sigma}\nonumber\label{H-MF},
\end{eqnarray}
where

\begin{eqnarray}
\Delta^{(1)}_{ij} &=&  - V_{0} \left( \langle  c_{j\downarrow} c_{i\uparrow} \rangle + \langle  c_{i\downarrow} c_{j\uparrow} \rangle \right)/2\\
W_{ij} &=& - \delta V_{ij} \langle c_{j\sigma}^\dagger c_{i\sigma} \rangle \\
\Delta^{(2)}_{ij} &=&  \delta V_{ij} \left( \langle c_{j\downarrow} c_{i\uparrow}  \rangle + \langle c_{i\downarrow} c_{j\uparrow}  \rangle \right)/2 \\
\mu_{ij\sigma} &=& \mu \delta_{ij} - V_{ij} \langle c_{j-\sigma}^\dagger c_{j-\sigma} \rangle
\end{eqnarray}
The pairing amplitude on a bond $(i,j)$, $\Delta^{(1)}_{ij}$, stems
from a constant nearest neighbor attraction, $-V_0$, and the density
wave pairings in the P-P and the P-H
channels are caused by a long-range interaction, $\delta V_{ij}$. Since our major interest is to study the orderings in
both channels at the same time, the chemical potential,
$\mu_{ij\sigma}$, is taken to be a constant.

The Bogoliubov-de Gennes equation is given by
\begin{eqnarray}
\left(
\begin{array}{cc}
\mathcal{H}_0 & \Delta^* \\
\Delta & -\mathcal{H}_0^*
\end{array}
\right)\left(
\begin{array}{c}
u_n(i)\\v_n(i)
\end{array}
\right)=E_n\left(
\begin{array}{c}
u_n(i)\\v_n(i)
\end{array}
\right)\label{bdg_ham}
\end{eqnarray}
where $\mathcal{H}_0$ and $\Delta$ are transfer matrices such that
\begin{eqnarray}
\mathcal{H}_0 x(i) &=&  -\left({\sum}_j (t + W_{ji})  + \mu_{ji}\right)x(j)\\
\Delta x(i) &=& {\sum}_j \left( \Delta^{(1)}_{ji} + \Delta^{(2)}_{ji}\right)     x(j)
\end{eqnarray}
where $x(i)$ can be either $u_n(i)$ or $v_n(i)$.

We numerically solve BdG eigenvalues $E_n$ and eigenvectors
$(u_n(i), v_n(i))$ on a lattice of $N$ sites with a periodic
boundary condition.  We then calculate the pairing amplitude, and
the density orderings in the P-P and the P-H
channels which are, respectively give by,

\begin{eqnarray}
\Delta_{ij}^{(1)} &=& -\frac{V_{0}}{2} \sum_n (u_n (j)
v^\ast_n(i) + u_n(i)v^\ast_n(j)) \\
\Delta_{ij}^{(2)} &=& \frac{\delta V_{ij}}{2} \sum_n (u_n (j)
v^\ast_n(i) + u_n(i)v^\ast_n(j)) \\
W_{ij} &=& -\delta V_{ij} \sum_n v_n
(j )v^\ast_n(i) .
\end{eqnarray}
 Substituting these equations back into the BdG equation, Eq.(\ref{bdg_ham}), and
repeating the same process until self-consisitency is achieved for each
of the local variables, we can obtain solutions of the BdG
equations. In the calculation, we have used Broyden's method for
efficient iteration.

\begin{figure*}
\includegraphics[width=17cm]{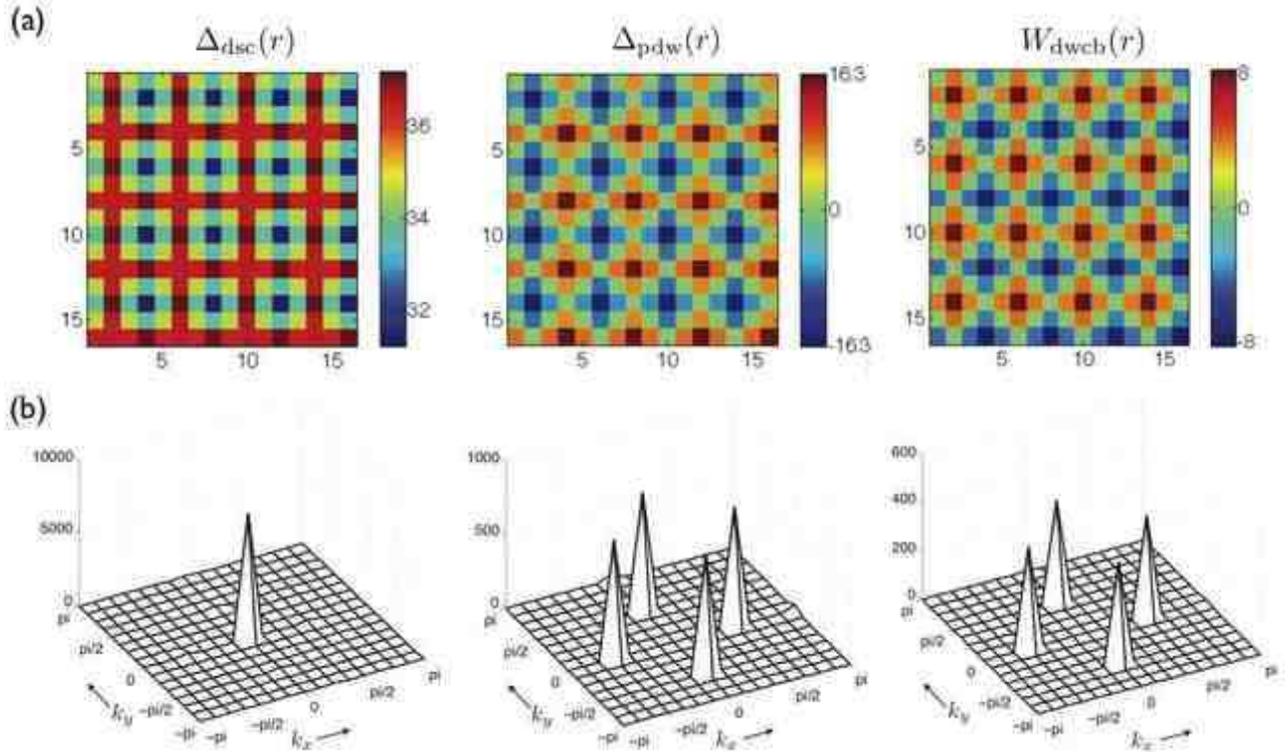}
\caption{
\label{fig_op_real_fourier_L16}
(a) The amplitude of the order parameters, $\Delta_{\text{DSC}}(r)$, $\Delta_{\text{PDW}}(r)$, and $W_{\text{DWCB}}(r)$ defined by Eq.~(\ref{dsc_r}), (\ref{pdw_r}), and (\ref{dwcb_r}), respectively,
plotted in 16$\times$16 lattice sites.
$\Delta_{\text{DSC}} (r) \approx \Delta_0$, $\Delta_{\text{PDW}} =  \Delta_1 \cos \mathbf{Q} \cdot r $,
and $W_{\text{DWCB}} (r)  = W_0 \cos \mathbf{Q} \cdot r $ with $\Delta_0 = 34.2$ meV, $\Delta_1 = 16.3$ meV, and $W_0 = 8.3$ meV, respectively.
$t = 125$ meV is chosen to have the amplitudes in units of meV.
(b) The Fourier transforms of the order parameters are displayed in the first Brillouin zone.
The prominent peaks are located at $\mathbf{Q} = \{ (\pm \pi/2,0), (0,\pm \pi/2)\}$ in both PDW and DWCB.
}
\end{figure*}

From the self-consistent solution, we can then compute the interesting order parameters. The $d$-wave pairing amplitude (DSC) at a lattice site $\mathbf{r}$ is determined by four nearest neighbors, $\Delta^{(1)}_{ij}$:
\begin{equation}
\Delta_{\text{DSC}}(\mathbf{r}) =  ( \Delta_{\mathbf{r},\mathbf{r}+\hat{x}}^{(1)} + \Delta_{\mathbf{r},\mathbf{r}-\hat{x}}^{(1)} - x \leftrightarrow y ) /4.\label{dsc_r}
\end{equation}
In the same way, an extended $s$-wave pairing density wave order in the P-P channels (PDW) can be obtained by $\Delta^{(2)}_{ij}$.
\begin{equation}
\Delta_{\text{PDW}}(\mathbf{r}) = ( \Delta_{\mathbf{r},\mathbf{r}+\hat{x}}^{(2)} + \Delta_{\mathbf{r},\mathbf{r}-\hat{x}}^{(2)}
+ x\leftrightarrow y  ) /4,\label{pdw_r}
\end{equation}
and $d$-wave density ordering in the P-H channel (DWCB) is
\begin{equation}
W_{\text{DWCB}}(\mathbf{r}) = ( W_{\mathbf{r},\mathbf{r}+\hat{x}} + W_{\mathbf{r},\mathbf{r}-\hat{x}} - x \leftrightarrow y  )/4.\label{dwcb_r}
\end{equation}

\begin{figure}
\includegraphics[width=9cm]{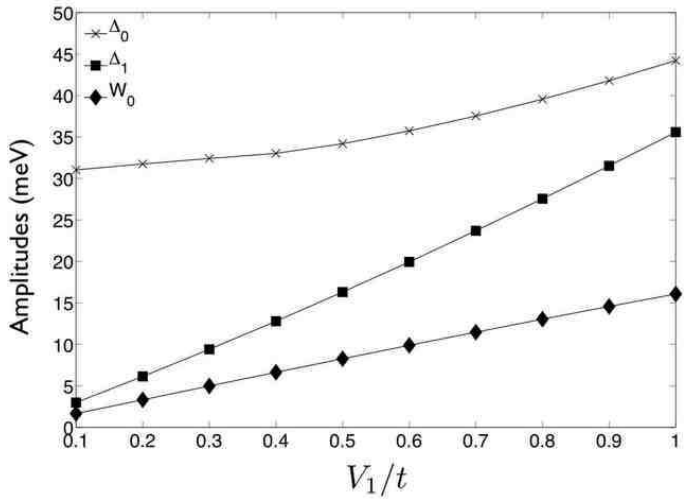}
\caption{
\label{fig_op_as_vnn}
The amplitudes, $\Delta_0$, $\Delta_1$, and $W_0$ of $\Delta_{\text{DSC}}(r)$, $\Delta_{\text{PDW}}(r)$, and $W_{\text{DWCB}}(r)$, respectively, as a function of the modulation long range interaction $V_1$. $t = 125$ meV is chosen to have the amplitudes in units of meV. For a $d$-wave superconducting order (DSC), $\Delta_0$, denoted by x along the curve for DSC, at each $V_1$ represents the mean value of $\Delta_{\text{DSC}}(r)$ which is slightly modulated. 
}
\end{figure}

We shall now present our results of full calculation. The uniform part of the interaction is set to $V_{0} = - 2.5t$ while the modulation part has amplitudes of $ V_1$ with the range of $0.1t$ to $1.0t$. The solutions we have obtained are independent of the initial guesses for the local variables. In Fig.~\ref{fig_op_real_fourier_L16} (a), the order parameters, $\Delta_{\text{DSC}}(\mathbf{r})$, $\Delta_{\text{PDW}}(\mathbf{r})$, and $W_{\text{DWCB}}(\mathbf{r})$, in a 2D real space are plotted for $V_0=2.5t$ and $V_1 = 0.5t$ with $t = 125$ meV. The P-H channel ordering, $W_{\mathbf{r},\mathbf{r}'}$ shows $d$-wave symmetry such that
\begin{eqnarray}\label{eqn:Wrr}
W_{\mathbf{r},\mathbf{r}'} = \left\{ \begin{array}{rl}
                       W_0 \left( \cos \mathbf{Q} \cdot \mathbf{r} + \cos \mathbf{Q} \cdot \mathbf{r}' \right) , & \mathbf{r}' = \mathbf{r} \pm \hat{x}\\
                       -W_0 \left( \cos \mathbf{Q} \cdot \mathbf{r} + \cos \mathbf{Q} \cdot \mathbf{r}' \right), & \mathbf{r}' = \mathbf{r} \pm \hat{y}
                       \end{array}
                       \right.
\end{eqnarray}
where $W_0 = 0.06t = 8.3$ meV and the order wavevectors $\mathbf{Q} = \{ (\pm \pi/2,0),(0,\pm \pi/2)\}$ as shown in Fig~\ref{fig_op_real_fourier_L16} (b).

The order in P-P channel has both $d$-wave, $\Delta^{(1)}_{\mathbf{r},\mathbf{r}'}$, and extended $s$-wave, $\Delta^{(2)}_{\mathbf{r},\mathbf{r}'}$, part. The extended $s$-wave part is a pure modulation described by
\begin{eqnarray}\label{eqn:PDWrr}
\Delta^{(2)}_{\mathbf{r},\mathbf{r}'} = \left\{ \begin{array}{rl}
                       \Delta_1 \left( \cos \mathbf{Q} \cdot \mathbf{r} + \cos \mathbf{Q} \cdot \mathbf{r}' \right) , & \mathbf{r}' = \mathbf{r} \pm \hat{x}\\
                       \Delta_1 \left( \cos \mathbf{Q} \cdot \mathbf{r} + \cos \mathbf{Q} \cdot \mathbf{r}' \right), & \mathbf{r}' = \mathbf{r} \pm \hat{y}
                       \end{array}
                       \right.
\end{eqnarray}
where $\Delta_1 = 0.13t = 16.3$ meV and the same order wavevectors $\mathbf{Q}$.
We have also obtained the results of $d$-wave pairing amplitude, or,$\Delta_{\mathbf{r},\mathbf{r}\pm\hat{x}}^{(1)}  = \Delta_0 $ and  $\Delta_{\mathbf{r},\mathbf{r}\pm\hat{y}}^{(1)}  = - \Delta_0 $ with $\Delta_0 = 0.28 t $.
In Fig.~\ref{fig_op_as_vnn} the modulation amplitudes of $\Delta_{\text{DSC}}(\mathbf{r})$, $\Delta_{\text{PDW}}(\mathbf{r})$, and $W_{\text{DWCB}}(\mathbf{r})$ were plotted as a function of $V_1$.

The self-consistent BdG calculation shows that there is indeed a
complementary connection between the density order in the
P-P channel (PDW) and in the P-H channel
(DWCB). Since PDW and DWCB are based on the same modulating
interaction, $\delta V_{ij}$, not only PDW and DWCB should be
present simultaneously  with $d$-wave superconductivity but also the
symmetries and order wavevector $\mathbf{Q}$ of both orders should be
closely related to each other. Therefore, the mean-field solutions
in the disordered $d$-wave superconductivity should include a
$d$-wave density wave with $\mathbf{Q}$ in the P-H pairing and an
extended $s$-wave density wave with the same $\mathbf{Q}$ in the
P-H pairing. Furthermore it is important to take both density
orders  into account on the same footing when we try to understand
the electronic states of the disordered high $T_c$ cuprates and the
pseudogap state at high temperature.


\section{The Mixed DWCB and PDW  state}

Analytically, the general features in STM measurements can be
captured by the DWCB. Due to the anisotropy inherited from the
$d$-wave factor of pairing, a weak DWCB order has a much stronger
effect on the antinodal region than on the nodal region. Thus it
naturally explains the puzzling dichotomy between the nodal and
antinodal excitations in high temperature superconductors. The local
phase fluctuations of Cooper pairs lead to a local modulation of
$d$-wave ordering in the P-H channel (DWCB), which
strongly affects the antinodal single particle excitations, as well
as an extended $s$-wave order in the P-P channel
(PDW).  Since both PDW and DWCB are bond-centered,
$\rho_{\mathbf{Q}}(\omega)$ will be an even function of
$\omega$~\cite{CHEN2004}. This symmetry distinguishes the PDW and
DWCB orders from the typical CDW in the P-H channel. The existing
experimental results are consistent with the even case.

The purpose of this section is to determine the symmetries of
density orders in P-P and P-H channels and then study the pseudogap physics under the presence of the orderings in P-P and P-H channels. Due to the complementary
connection, if the PDW is an extended $s$-wave, DWCB is a $d$-wave
P-H pairing. If the PDW is a $d$-wave, the P-H
pairing density order must be extended $s$-wave. It is worth noting
that the symmetries of the orders are very important to
  explain the behavior of the conductance spectra at
low energy.  We will show that in DSC state the $d$-wave
P-H pairing(DWCB) and the extended $s$-wave
P-P pairing(PDW) can explain the low energy spectra
while the solution with other symmetries cannot sufficiently
match the experimental results.

In section II, we have shown that, even without PDW, the presence of DWCB and DSC orders in a pseudogap state can
capture the features of the Fermi arc. To understand the temperature dependence of the length of the Fermi arc above $T_c$ we assumed the temperature dependence of the imaginary part of the energy $\eta$  in the Green function implicitly.
Using the results of the self-consistent calculations, this assumption of the temperature dependence of $\eta$ is not necessary. The presence of the complementary connected $d$-wave orders in P-P and the extended $s$-wave order in P-H channels with DSC order naturally lead us to the temperature
dependence of the length of the Fermi arc within the Franz and Millis model~\cite{FRANZ1998}.

\subsection{The symmetries in P-P and P-H channels}

In this subsection we study how the local density of states is associated with the symmetries of the complementary
connected orders in P-P and P-H channels in a disordered $d$-wave superconducting state.

The local density of states(LDOS) in the DSC state in the presence
of P-P and P-H channel orders  is calculated by
\begin{equation}
\rho(\mathbf{r},\omega) = \sum_n [|u_n(\mathbf{r})|^2 \delta(\omega-E_n) + |v_n(\mathbf{r})|^2 \delta(\omega + E_n) ]
\end{equation}
The averaged DOS is given by
\begin{equation}
\rho(\omega) = \sum_\mathbf{r} \rho(\mathbf{r},\omega)
\end{equation}
and the Fourier components at the wavevectors $\mathbf{Q} = \{
(\pi/2,0),(0,\pi/2)\}$ by
\begin{equation}
\rho_{\mathbf{Q}} (\omega) = \sum_\mathbf{r} e^{i\mathbf{Q}\cdot r} \rho(\mathbf{r},\omega)
\end{equation}
We have calculated them  in a simple band structure, $t =
-125\text{meV}$, and $\mu = 0$.

\begin{figure}
\includegraphics[width=8cm]{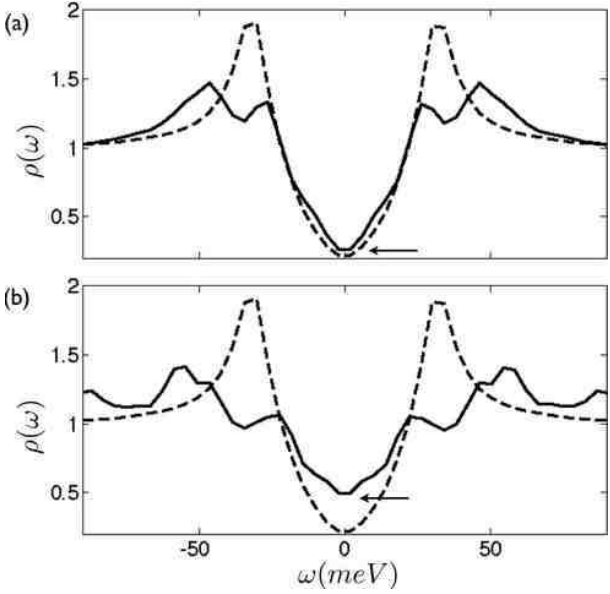}
\caption{
\label{fig_dos_dsc_pdw_dwcb}
(a) The averaged DOS are plotted for $\Delta_{\text{DSC}} = 33$ meV in the case of $d$-wave P-H and extended $s$-wave P-P channel orders: $\Delta^{}_{\text{PDW}} = 12.7$ meV, $W^{}_{\text{DWCB}} = 3.3$ meV. The arrow indicates $\delta h$, that is, DOS at $\omega = 0$ off of one for pure $d$-wave superconducting state. $\delta h = 0.0599$.
(b) The averaged DOS in the mixed state of DSC ($\Delta_{\text{DSC}}=33$ meV) with $d$-wave P-P and extended $s$-wave P-H channel orders. Here we plotted one with $\Delta^{}_{\text{PDW}} = 17.3$ meV, and $W^{}_{\text{SWCB}} = 10$ meV. $\delta h = 0.2933$.}
\end{figure}

For the $d$-wave P-H and extended $s$-wave P-P pairing, we have used $V_0 = 2.5t$ and various $V_1 $  as in the table~(\ref{table:dwcb}), where order parameters as solutions of the BdG calculation are given in meV, and $\delta h$ is the height of $\rho(\omega)$ at $\omega =0$ off of the superconducting DOS displayed in a dashed line in Fig.~\ref{fig_dos_dsc_pdw_dwcb}.
\begin{equation}
\label{table:dwcb}
\begin{array}{c|ccccc}\hline
V_1 & \Delta_{\text{DSC}} &  \Delta_{\text{PDW}}  & W_{\text{DWCB}} &&    \delta h  \\\hline
0.4t & 33.0 & 12.7 & 3.3  & &   0.0599  \\
0.6t & 35.7 & 19.9 & 9.9 &   &  0.1125 \\
0.8t & 39.6 & 27.5 & 13.0  &  &  0.1445  \\
1.0t & 44.2 & 35.6 & 16.1  &   & 0.1546  \\
\hline
\end{array}
\end{equation}
For the extended $s$-wave P-H and $d$-wave P-P channel orders, we have used $V_0 = 2.5t$ and $V_1 $ as in~(\ref{table:swcb}), where $W_{\text{SWCB}}$ denotes the extended $s$-wave P-H pairing for  convenience.
\begin{equation}
\label{table:swcb}
\begin{array}{c|ccccc}\hline
V_1 &  \Delta_{\text{DSC}}    &   \Delta_{\text{PDW}}   &  W_{\text{SWCB}} & &  \delta h   \\\hline
0.4t  &  31.5&    11.5&      6.6& &   0.1144\\
0.6t &  32.7&    17.3&      10.0&  &  0.2933\\
0.8t &  34.2&    22.9&       13.3 & & 0.4200\\
1.0t &  36.2&    28.4&      16.6&    &0.3589\\
\hline
\end{array}
\end{equation}

The BdG calculation shows that DWCB, which is $d$-wave P-H pairing
order, is crucial to the physics in the disordered
superconductor. It captures the experimental observation of
conductance spectrum at both low and high energy. As expected, the
$d$-wave symmetry of the P-H pairing density order has little effect
at low energy while it strongly affects the superconducting
coherence peaks. Therefore we can conclude the competing order in
the P-H channel must have $d$-wave symmetry. Due to the
complementary connection in the orders in P-H and P-P channels, it
naturally lead to the existence of the pair density order that has
extended $s$-wave symmetry.

In Fig.~\ref{fig_dos_dsc_pdw_dwcb}, to illustrate the above argument, we plotted the averaged DOS in two cases: (a) $d$-wave P-H and extended $s$-wave P-P channel orders, and (b) $s$-wave P-H and extended $d$-wave P-P channel orders in the $d$-wave superconducting state ($\Delta_{\text{DSC}} = 33\text{meV}$).

\subsection{The DWCB and PDW in the pseudogap state}

Now let us consider the spectral function $A(\mathbf{k},\omega)$ in the pseudogap state to capture the features in the Fermi arc. First we rewrite the mean-field Hamiltonian for PDW and DWCB in momentum space. Taking the Fourier transform of Eq.~(\ref{eqn:mfh}), we obtain
\begin{eqnarray}
\label{eqn:pdwh}
\mathcal{H}_{\text{PDW}} &=& \sum_{\mathbf{r},\delta,\sigma}  \Delta^{(2)}_{\mathbf{r},\mathbf{r}+\delta} c_{\mathbf{r},\sigma}^\dagger c_{\mathbf{r}+\delta,-\sigma}^\dagger + \text{h.c.}  \\
&=&\sum_{\mathbf{p},\mathbf{q},\sigma}\left[ \sum_{\mathbf{r},\delta}  \Delta_{\mathbf{r},\mathbf{r}+\delta}^{(2)} e^{-i ( \mathbf{p} + \mathbf{q} )\cdot\mathbf{r}}e^{-i\mathbf{q}\cdot\delta}\right] c_{\mathbf{p},\sigma}^\dagger c_{\mathbf{q},-\sigma}^\dagger ,\nonumber
\end{eqnarray}
where $\delta$ denotes the unit vectors for the nearest-neighbors of $\mathbf{r}$.
Since $\Delta^{(2)}_{\mathbf{r},\mathbf{r}+\delta} = \Delta_1/2 \sum_{\mathbf{Q}} [\cos \mathbf{Q}\cdot \mathbf{r} + \cos \mathbf{Q}\cdot (\mathbf{r}+\delta) ]$, the term in the square bracket in Eq.~(\ref{eqn:pdwh}) is rewritten as
\begin{eqnarray*}
&&\Delta_1 \sum_{\mathbf{r},\mathbf{Q},\delta}[ \cos \mathbf{Q}\cdot \mathbf{r} + \cos \mathbf{Q}\cdot (\mathbf{r} +\delta) ] e^{-i(\mathbf{p} + \mathbf{q})\cdot \mathbf{r} } e^{-i\mathbf{q}\cdot \delta} \\
 = &&\Delta_1 \sum_{\mathbf{Q},\delta} \left( e^{-i\mathbf{q} \cdot\delta} + e^{i\mathbf{p}\cdot\delta} \right)  \delta_{\mathbf{p} + \mathbf{q},\mathbf{Q}} \\
 =&& \Delta_1 \sum_\mathbf{Q} [ \cos p_x + \cos (p_x + Q_x ) + x \leftrightarrow y ]\delta_{\mathbf{p} + \mathbf{q}, \mathbf{Q}}
 \end{eqnarray*}
Thus $\mathcal{H}_{\text{PDW}}$ in momentum space can be given by
\begin{equation}
\mathcal{H}_{\text{PDW}} = \sum_{\mathbf{k},\mathbf{Q},\sigma} \Delta_\mathbf{k}^{(2)} c_{\mathbf{k},\sigma}^\dagger c_{-\mathbf{k}-\mathbf{Q},-\sigma}^\dagger + \text{h.c.}
\end{equation}
where $\Delta^{(2)}_\mathbf{k} = \Delta_1 [ \cos k_x + \cos \left( k_x + Q_x \right) + x \leftrightarrow y ]$.

The momentum space expressions for DWCB in $\mathcal{H}_{\text{MF}}$, Eq.~(\ref{eqn:mfh}), can be obtained in the same way:
\begin{equation}
\mathcal{H}_{\text{DWCB}} = \sum_{\mathbf{k},\mathbf{Q},\sigma} W_\mathbf{k} c_{\mathbf{k}+\mathbf{Q},\sigma}^\dagger c_{\mathbf{k},\sigma} + \text{h.c.}
\end{equation}
where $ W_\mathbf{k} = W_0 [ \cos k_x + \cos (k_x + Q_x ) - x \leftrightarrow y ]$.
Therefore the mean-field Hamiltonian, Eq.~(\ref{eqn:mfh}), can be recast in momentum space as
\begin{eqnarray}
\mathcal{H}_{\text{MF}} &=& \sum_{\mathbf{k},\sigma} \left( \xi_\mathbf{k} c{\mathbf{k},\sigma}^\dagger c_{\mathbf{k},\sigma} + \Delta^{(1)}_\mathbf{k} c_{\mathbf{k},\sigma}^\dagger c_{-\mathbf{k},-\sigma}^\dagger \right)  \\
&+& \sum_{\mathbf{k},\mathbf{Q},\sigma}\left( \Delta^{(2)}_\mathbf{k} c_{\mathbf{k},\sigma}^\dagger c_{-\mathbf{k}-\mathbf{Q},-\sigma}^\dagger + W_\mathbf{k}\, c_{\mathbf{k}+\mathbf{Q},\sigma}^\dagger c_{\mathbf{k},\sigma}\right) + \text{h.c.} \nonumber
\end{eqnarray}
where $\xi_\mathbf{k} = -t/2 ( \cos k_x + \cos k_y ) - \mu$, and $ \Delta^{(1)}_\mathbf{k} = \Delta_0/2 (\cos k_x - \cos k_y)$.
In the same way as in section II, it can be re-expressed in terms of the Nambu formalism as such
\begin{equation}
\mathcal{H}_{\text{MF}} = \sum_\mathbf{k} \psi_\mathbf{k}^\dagger H(\mathbf{k}) \psi_\mathbf{k}
\end{equation}
where $\psi_\mathbf{k} = \left( c_{\mathbf{k}\uparrow}, c_{\mathbf{k}+\mathbf{Q}\uparrow}, c_{-\mathbf{k}\downarrow}^\dagger, c_{-\mathbf{k}-\mathbf{Q}\downarrow}^\dagger \right)^\dagger$, and
\begin{equation}
H(\mathbf{k}) = \left( \begin{array}{cccc}
\xi_\mathbf{k} & W_\mathbf{k} & \Delta^{(1)}_\mathbf{k} & \Delta^{(2)}_\mathbf{k} \\
W_\mathbf{k}^\ast & \xi_{\mathbf{k}+\mathbf{Q}} & \Delta^{(2)}_{\mathbf{k}+\mathbf{Q}} & \Delta^{(1)}_{\mathbf{k}+\mathbf{Q}} \\
\Delta_\mathbf{k}^{(1)\ast} & \Delta^{(2)\ast}_{\mathbf{k}+\mathbf{Q}} & \xi_{-\mathbf{k}} & W_{\mathbf{k}+\mathbf{Q}}^\ast \\
\Delta^{(2)\ast}_{\mathbf{k}} & \Delta^{(1)\ast}_{\mathbf{k}+\mathbf{Q}} & W_{\mathbf{k}+\mathbf{Q}} & \xi_{-\mathbf{k} + \mathbf{Q}}
\end{array}
\right)
\end{equation}

The spectral function is given by the imaginary part of the Green function,
\begin{equation}
G_{11}(\mathbf{k},\omega) = \langle \Omega| [ \omega + i \eta - H(\mathbf{k}) ]_{11} ^{-1} |\Omega \rangle
\end{equation}
The ground state $|\Omega\rangle$ is defined by the state without quasiparticles. In the pseudogap phase, it can be calculated within the model of Franz and Millis~\cite{FRANZ1998}:
\begin{equation}
A(\mathbf{k},\omega) = -\frac{1}{\pi} \int d\omega' P(\omega') \text{Im} G_{11}(\mathbf{k},\omega - \omega'),
\end{equation}
where $\omega'$ is the energy shift due to the presence of a uniform supercurrent, and $P(\omega')$ is the probability distribution of $\omega'$ calculated within a $2D$ $XY$ model as such
\begin{equation}
P(\omega') = \sqrt{2\pi} W(T) e^{-\omega'^2/2W^2(T)}
\end{equation}
where $W(T) $ is an order of pseudogap and an increasing function of temperature above $T_c$.

\begin{figure}
\includegraphics[width=8cm]{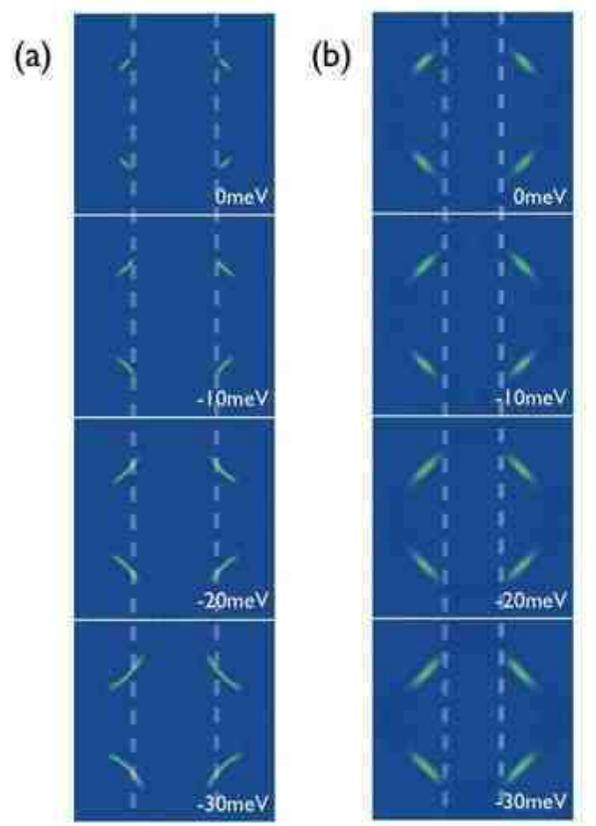}
\caption{
\label{fig_Akw_PG_as_E}
The energy dependence of the the Femi arc.
(a) shows the development of the gapless region as a function of energy $\omega$ in the superconducting state without DWCB.
As $\omega$ increases(from top to bottom), the scattering wavevectors connecting one tip of the green curve to another tip decrease. However, in (b) the gapless region remains unchanged as $\omega$ increases, that is, the Fermi arc is non-dispersive.  }
\end{figure}

Now we will study the energy dependence of the Fermi arc in the
pseudogap state. One of the salient features of the pseudogap phase
is that the Fermi arc is non-dispersive, whereas it is dispersive in
the superconducting phase. The Fermi arc along the Fermi surface is
identified by the peak of EDC at $\omega = 0$.
Fig.~\ref{fig_Akw_PG_as_E}(b) demonstrates that the Fermi arc
in green remains unchanged as $\omega$ is
increased. In the superconducting state, however, the gapless region
develops as the energy decreases below the Fermi level as seen in
Fig.~\ref{fig_Akw_PG_as_E}(a).

\begin{figure*}
\includegraphics[width=17cm]{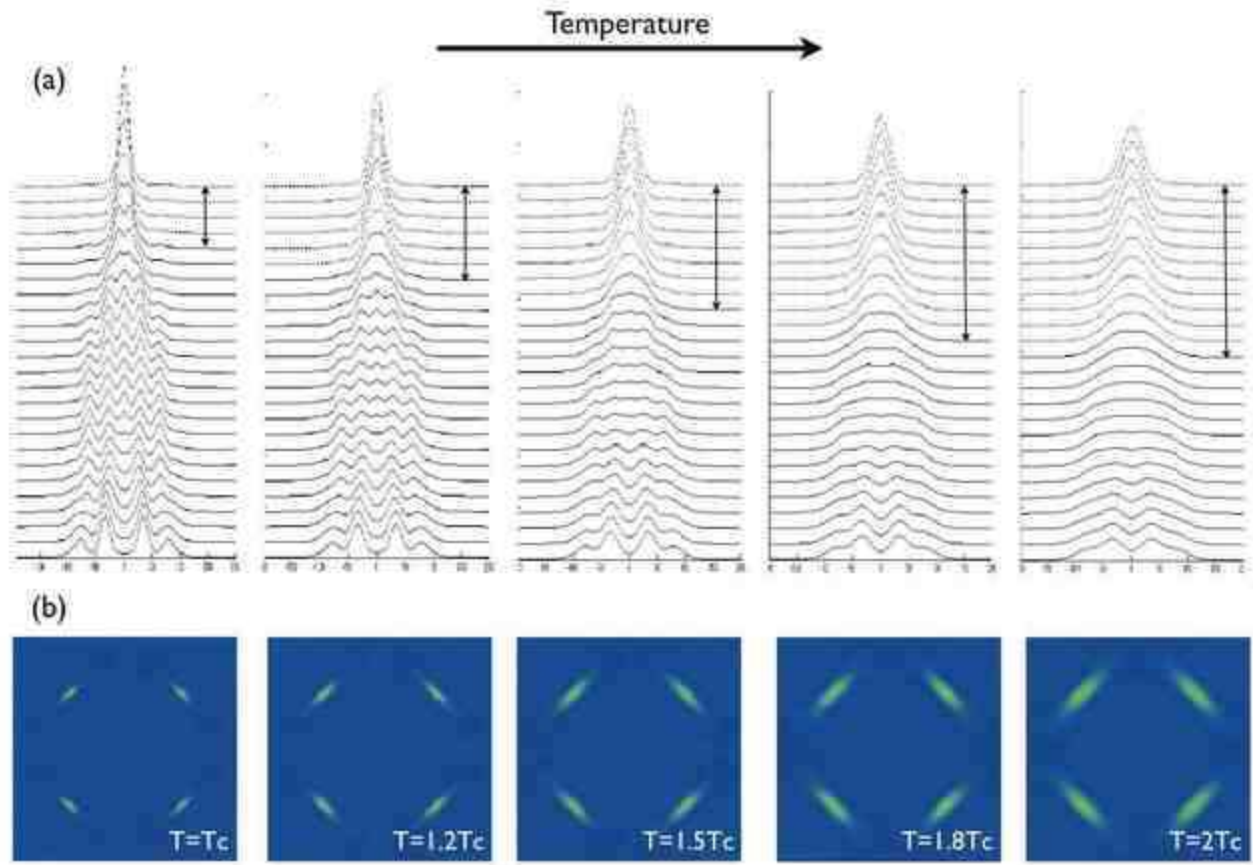}
\caption{
\label{fig_Akw_PG_as_T}
The temperature dependence of the Femi arc above $T_c$.
(a) EDC's along the Fermi surface from the nodal point (top) to anti-nodal point (bottom) are plotted.
The arrows indicate the gapless part in the Fermi surface and show the length of the Fermi arc linearly depends on the temperature in the pseudogap state.
(b) The Fermi arc as a function of the temperature $T$ at above $T_c$.}
\end{figure*}

While the Fermi arc is non-dispersive, it is found that the length of it is increasing linearly as a function of temperature~\cite{Kanigel2006}. Within the model of Franz and Millis, when $T> T_c$, $W(T)$ is an approximately linearly increasing function when the order parameters remain fixed. Adopting the values of $W(T)$ above $T_c$, we have calculated $A(\mathbf{k},\omega)$ as a function of temperature.

In Fig.~\ref{fig_Akw_PG_as_T}(a), EDC are plotted from the nodal to
antinodal point along the Fermi surface. Clearly it shows a gapless
nodal point and gapped anti-nodal point as indicated by darker
curves. But, contrary to a superconducting state where the Fermi surface
is gapped except at the nodal point, there is a region, the so called
Fermi arc, where peaks still survive at $\omega=0$ and thus the superconducting gap is closed. The curve on top is the EDC at the
nodal point, $(\pi/2,\pi/2)$, perpendicular to the Fermi
surface. It is manifest that the size of the arrows is increasing as
the temperature rises. Fig.~\ref{fig_Akw_PG_as_T}(b) shows the Fermi
arc in the first Brillouin zone. As in the
Fig.~\ref{fig_Akw_PG_as_E} the green color represents the gapless
region of the Fermi surface, that is, the Fermi arc.


\section{CONCLUSION AND DISCUSSION}

We have presented a detailed analysis of  competing orders in cuprates
which  explains the checkerboard pattern observed in STM spectra. We
have shown that in general there is a complementary connection
between the orders  in the P-P and the P-H
channels in both the superconducting state and the pseudogap state.
The symmetries of the orders in both channels are closely related to
each other. In the $d$-wave superconducting state, the presence of
the $d$-wave(extended $s$-wave) P-H density order
implies the coexistence of the extended $s$-wave($d$-wave)
P-P pairing order. The self-consistent calculations in
the disordered superconducting state result in solutions with
uniform $d$-wave superconducting order,  a P-H channel
checkerboard density order with the order wavevector  $\mathbf{Q}$, and a
P-P channel pairing density wave with the same
$\mathbf{Q}$. The symmetries of the channel orders is crucial to the
effects on the density of states in STM experiments. We have
found that $d$-wave checkerboard density order in the
P-P and the extended $s$-wave pair density wave are in
good agreement with the experimental data in the superconducting
phase.

The coexistence of orders in both P-P and P-H channels are important
to understand the pseudogap phase which can be considered as a
state which maintains pairing amplitude without phase coherence.
The fluctuation of phase would naturally lead to  the DWCB and PDW
orders. We have shown that the presence of DWCB and PDW results in
a non-dispersive Fermi arc and  our calculations show  the linear
dependence of the length of the Fermi arc at temperatures above
$T_c$.  Moreover, the effect of the presence of PDW and DWCB  on
single particle spectra is much larger in the antinodal direction
than in the nodal direction. These results  also explain the
dichotomy between the nodal and antinodal single particle spectra in
the cuprates.

The connections of the P-P and P-H channel orders  are important in
formulating the effective low energy theory in cuprates. From our
results in this paper, the low energy effective theory should
include both orders. So far, most theories have treated them
independently. It is also clear that the presence of both orders can
result in new physics in transport and thermal properties. A
detailed study of these effects will be reported elsewhere.

{\it Acknowledgement:}   J. P. Hu and K. Seo  are supported by
National Scientific Foundation under award number Phy-0603759. H.D.
Chen  is supported by the U.S. Department of Energy, Division of
Materials Sciences under Award No. DEFG02-91ER45439, through the
Frederick Seitz Materials Research Laboratory at the University of
Illinois at Urbana-Champaign.

\bibliography{DWCB}

\end{document}